\documentclass[12pt]{article}
\usepackage{amssymb,epsfig}

\begin{document}
\begin{center}
\Large
Explosive Percolation in Social and Physical Networks\\
\end{center}

\begin{center}
Eric J. Friedman$^{1,2}$ and  Joel Nishimura$^2$\\$ $ \\
$^1$School of ORIE \\
and\\
$^2$Center for Applied Mathematics\\
Cornell University \\
Ithaca, NY, 14850\\
\verb+{ejf27,jdn48}@cornell.edu+

First Version: December 1, 2009

Current Version: January 26, 2009

\end{center}

\section{Introduction}

Recently Achlioptas, D'Souza and Spencer \cite{ADS09} uncovered a surprising property in some random networks.  They showed that for several Achlioptas Processes the (normalized) growth of the giant component is discontinuous in large networks, as seen in Figure~1. (Note that this does not necessarily correspond to a first order phase transition, e.g. \cite{Zif09}). This has the disconcerting implication that giant clusters appear (essentially) without warning; a property which could have important implications in many settings.  However, the Achlioptas processes are difficult to interpret and analyze, so their results are based on careful (and clever) numerical analysis, but lack formal proofs.

\begin{figure}
	\centering
		\epsfig{file=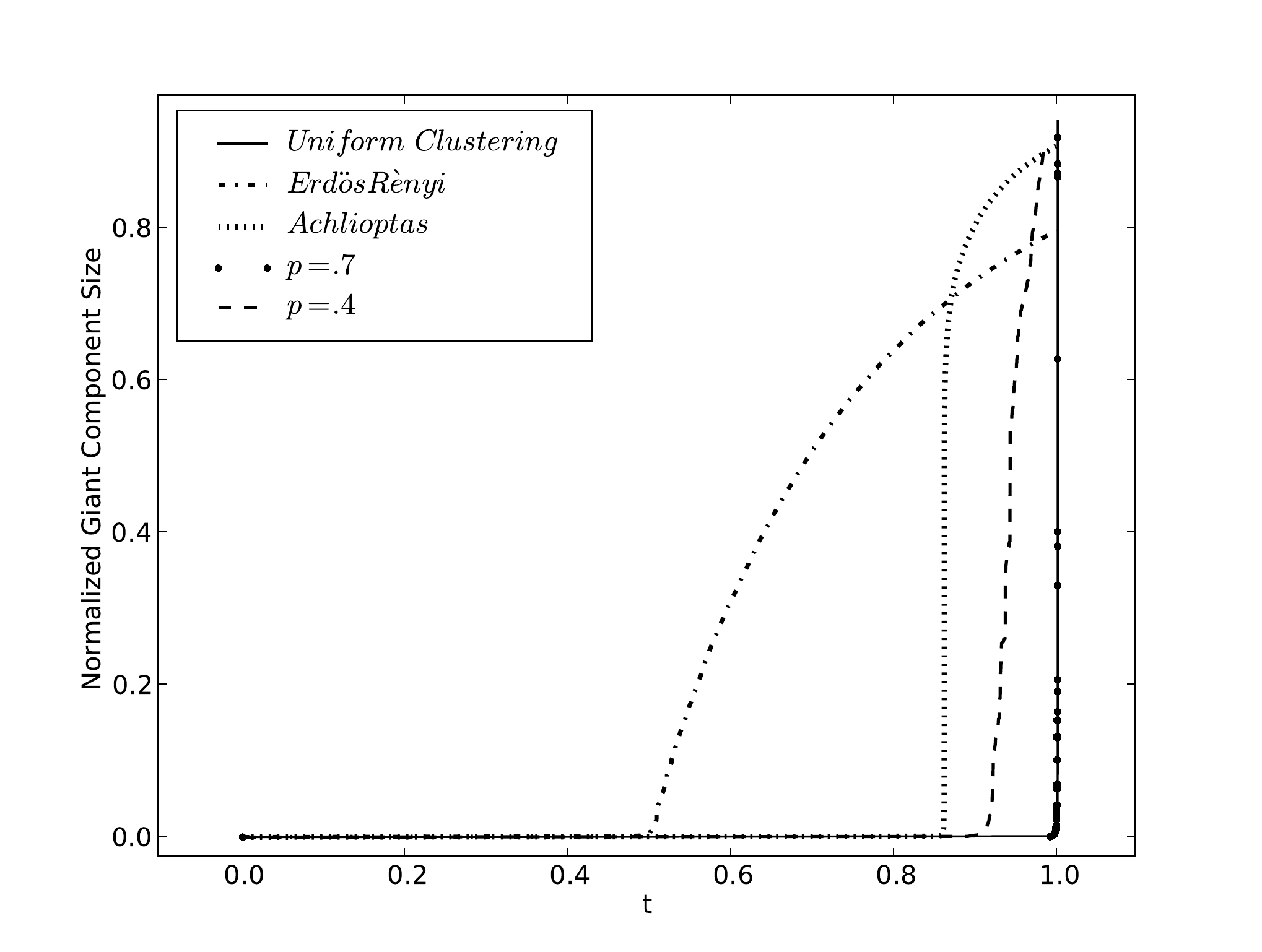,width=0.7\linewidth,clip=}
		\caption{The transition diagrams for the Erd\H os  R\'enyi, Achlioptas sum rule, weighted social networks ($p=.5$ and $p=.7$) and uniform clustering  models (p=0). Note that all except the Erd\H os  R\'enyi are explosive.}
\end{figure}

In this brevium, we extend their work by presenting several random network models which are quite simple and have natural interpretations as biological, physical and social networks. In addition, they are straightforward to analyze and have straightforward proofs. (Such proofs are important, as the numerical analysis is delicate and can be misleading.)

\section{Simple Models}

Our first model can be viewed as a model of social networks in which nodes represent people and edges represent social connections. In the classic Erd\H os  R\'enyi model \cite{ErR60} it is assumed that connections are formed randomly with each person having the same probability of making a connection with any other. However, in our model we assume that people who are already part of a large social network have less incentive to forge new connections. To formalize this we assume that the probability of choosing any person in a new connection is inversely proportional to the size of their current social network raised to the $p$'th power.  For $p=0$ this is the Erd\H os  R\'enyi model and has a continuous transition; however, for $p=1$ this has an explosive transition, as we prove below (and show in Figure~1).  Interestingly, for all $p>1/2$ numerical experiments suggest that the the transition remains a step function at $t=1$ while for $0<p<1/2$ it appears to remain explosive. (See Figure~1.) 

Our second model is in the class of preferential attachment models which have been used in many applications, especially models of the Internet \cite{BaA99}. In this rather extreme model we assume that edges are chosen at random between nodes of degree less than 2.  This corresponds to a social network where no person has more than 2 connections. 

Our final model is extremely simple and forms the basis for our analysis.  In this ``uniform clustering''  model  we simply ``merge'' random clusters, chosen uniformly. For example  we can start with clusters $\{1\}\{2\}\{3\}\{4\}$
then merge 2 at random to get clusters $\{1,3\}\{2\}\{4\}$. Then another random merge to get clusters $\{1,3,4\}\{2\}$ etc.  This model appears to have many applications in both social and physical networks and is equivalent to Kingman's analysis of coalescents in genetics \cite{Kin82}.

\section{Proving Explosiveness}

To show that the uniform clustering model has an explosive transition we apply the ideas introduced by Friedman and Landsberg in \cite{FrL09}. Consider the distribution of cluster sizes after $n-n^{1/2}-1$ merges, when there will be exactly $n^{1/2}$ clusters remaining.  As shown in \cite{FrL09}, if this distribution is sufficiently uniform then an explosive transition will occur.

Kingman's analysis \cite{Kin82} shows that all clusterings are equally likely. Using this it is easy to show that the probability of the largest chain being longer than $n^{3/4}$ is less than $n^{1/2}e^{-n^{1/4}}$ which vanishes as $n$ goes to infinity. Thus, the width of the transition region in the rescaled graph is $n^{1/2}/n$ so in the limit the graph of the transition function is simply a step function at 1.

This model is essentially the same as our first model with $p=1$ and can be shown to have the same statistical properties as our second, extending our proof of explosiveness to them.

\section{Concluding Comments}

Our analysis suggests that explosive transitions might be much more common than typically expected in biological, social and physical networks.

\paragraph*{Acknowledgment} After initial submission of this paper we became aware of
\cite{MaChat} and \cite{K09}, which both explore models similar to
ours.  Specifically \cite{MaChat} uses numerical evidence to look at
the behavior for different $p$ values on the random graph and the grid
while \cite{K09} connects the model to the Smoluchowski coagulation
equation to provide a non combinatorial proof as well as other
insights.

The authors thank Rick Durrett for pointing out the connections to coalescents, Adam Landsberg, Seth Marvel, and Steve Strogatz for helpful suggestions. This research has been supported in part by the NSF under grant  CDI-0835706.


\end{document}